# Limitation of simple np-n tunnel junction based LEDs grown by MOVPE

Y. Robin[1,*], Q. Bournet[2], G. Avit[1], M. Pristovsek[1], Y. André[3], A. Trassoudaine[3] and H. Amano[1]

[1] *Institute of Materials and Systems for Sustainability (IMaSS), Nagoya University, Japan*
[2] *Polytech Clermont-Ferrand, Université Clermont Auvergne, France*
[3] *Institut Pascal, Université Clermont Auvergne, France*

------------------------

[*] *Yoann.robin@crhea.cnrs.fr*

We show evidence that tunnel junctions (TJs) in GaN grown by metal-organic vapor phase epitaxy are dominated by trap-assisted (Poole-Frenkel) tunneling. This stems from observations of the careful optimized doping for the TJs. Especially the $p^{++}$ and the $n^{++}$ layers are far from ideal. The $n^{++}$ layer induces 3D growth, which can be seen by a rising oxygen signal in Secondary Ions Mass Spectroscopy (SIMS). Furthermore, Mg segregation observed by SIMS indicates a depletion region of more than 10 nm. Still, we could realize TJ based LEDs with a low penalty voltage of 1.1 V and a specific differential resistance of about $10^{-2}\,\Omega.cm^2$ at 20 mA without using an InGaN interlayer.

## I. INTRODUCTION

Realization of high quality p-n homojunction based on wide bandgap semiconductors is challenging due to doping asymmetry problems [1]. GaN, with a bandgap of about 3.4 eV, does not deviate the rule. While doping n-type is rather simple, achieving decent p-type layers is not straightforward [2,3].

Intentional n-type doping of GaN is traditionally realized by substituting gallium by silicon, which allows to reach electron concentration up to $10^{19}\,cm^{-3}$. However, at higher doping levels, Si doped GaN layers tend to roughen or crack under tensile stress [4,5]. Recently, germanium, whose size is comparable to gallium, has been studied as alternative dopant to cover the $10^{19}$-$10^{20}\,cm^{-3}$ range [6-8]. Oxygen on N-site is an efficient n-type donor, although only few reports exist on intentional oxygen doping [9-11]. N-type doping of GaN by other chalcogens (sulfur, selenium…) substituting the nitrogen site is little investigated. However their huge size difference to nitrogen does not bode well for achieving high doping levels and the relatively low vapor pressures of these solid easily cause strong carry-over [12,13].

For p-doped GaN, there are even less suitable acceptors. Zinc, a standard dopant in other III-V materials, is still a relatively deep acceptor in GaN ($E_a$~300-400 meV) [14]. Carbon exhibits an amphoteric behavior and beryllium is a shallow acceptor ($E_a$~100-140 meV) but suffers from severe self-compensation [15-18]. Other unsuccessful attempts were made using calcium, cadmium and mercury [19]. In the end, magnesium is the only dopant that allows a decent p-type conduction after an activation. Unfortunately, the maximum hole concentration barely reaches $10^{18}\,cm^{-3}$ due to the large ionization energy ($E_a$~150-200 meV) even tough mid $10^{19}\,cm^{-3}$ magnesium can be incorporated before onset of compensating defects [20].

Recently, tunnel junctions (TJs) have gained interest in nitrides. Such TJs could alleviate non-equilibrium hole injection, reduced optical loss in UV-LEDs, better current spreading in higher mobility n-type layers, inter-cavity contact in vertical cavity surface emitting lasers (VCSELs), or simpler vertical integration of tricolor micro-LEDs for micro-displays [21-26]. Esaki junctions typically require both low or moderate bandgap materials and doping levels around $10^{20}$ cm$^{-3}$ that is out of reach for GaN [27]. Nevertheless, GaN based TJ based LEDs reported surprisingly low penalty voltages [28-34]. The present paper aims to shed light on the efficiency of GaN based TJ grown by metal-organic vapor phase epitaxy (MOVPE). Hence, we first analyzed the crystal properties of p- and n-GaN layers separately. Then, we used them in simple np-n structure and corresponding TJ based LEDs. Based on those findings we discuss the tunneling mechanism.

## II. EXPERIMENTAL DETAILS

All the samples were prepared in a 3x2" EpiQuest vertical showerhead MOVPE reactor. For Si and Mg doping, pure $H_2$ was used as carrier gas. Both Mg:GaN and Si:GaN layers were grown using a V/III ratio of 7200, at a pressure of 150 Torr and at temperatures ranging from 975 to 1100°C with a growth rate of about 500 nm/h. Bisethylcyclopentadienyl-magnesium (EtCp$_2$Mg), tetramethyl-silane (TMSi), trimethyl-gallium (TMG) and ammonia (NH$_3$) were used as metal-organic precursors. For the growth of the quantum wells (QWs), the carrier gas was switched to pure $N_2$ and triethyl-gallium (TEGa) and trimethyl-indium (TMIn) were used as precursors.

Electrical characterizations were performed by Hall effect measurements in Van-der-Pauw geometry. For Si doping levels below $10^{19}$ cm$^{-3}$ Hall effect measurements and secondary ion mass spectroscopy (SIMS) concentrations agreed. Above that limit, the dopant concentration was measured by SIMS or given as extrapolation of our calibration. Photoluminescence (PL) was performed at room temperature (RT) using a He-Cd laser ($\lambda_{ex}$=325 nm, P=1 mW). For Mg:GaN layer activation, the samples were annealed at 700°C for 5 minutes under $N_2$. Surface morphologies were investigated by a Nanocute Hitachi atomic force microscope (AFM) working in tapping mode and by Nomarski optical microscopy (OM). Scanning transmission electron microscopy (STEM) pictures were recorded using a SU-70 Hitachi microscope with an operating voltage of 5 kV.

TJ based LEDs were processed by standard photolithography. L-shape mesa of 0.0675 mm$^2$ were prepared using Cl$_2$ dry etching. Then, the samples were annealed for 30 min at 750°C in $N_2$ ambient. Finally, Ti/Al electrodes were simultaneously deposited by e-beam lithography on both n-GaN terminals and annealed 5 min at 625°C in pure $N_2$. For the reference LED, p-GaN top layer was contacted by a Ni/Au electrode annealed 10 min in air at 500°C.

## III. RESULTS AND DISCUSSION

Fig. 1 summarizes the morphological, optical and electrical properties of 500 nm thick Mg doped GaN layers grown at various Mg/Ga ratios, which reflects current literature. The near band edge (NBE) emission of the PL slowly decreases and a characteristic donor-acceptor pair (DAP) related signal appears while increasing the Mg content [35]. A strong blue-band (BB) dominates the PL spectrum when the hole concentration reaches its maximum around few $10^{17}$ cm$^{-3}$ at RT, corresponding to a Mg concentration of few $10^{19}$ cm$^{-3}$ [36]. Further increase of the Mg content leads to the apparition of various compensating defects and pyramidal inversion domains (IDs) [37]. Simultaneously, the doping level drastically decreases and the p-GaN layers become resistive. At a Mg concentration of a few $10^{20}$ cm$^{-3}$ - which is the doping level typically required for TJs - the surface is rough and decorated with large hexagonal hillocks from N-polar inversion domains (IDs) likely induced by the formation of Mg$_3$N$_2$ sub-structures [38,39]. A high density of defects is indicated by the absence of any PL signal.

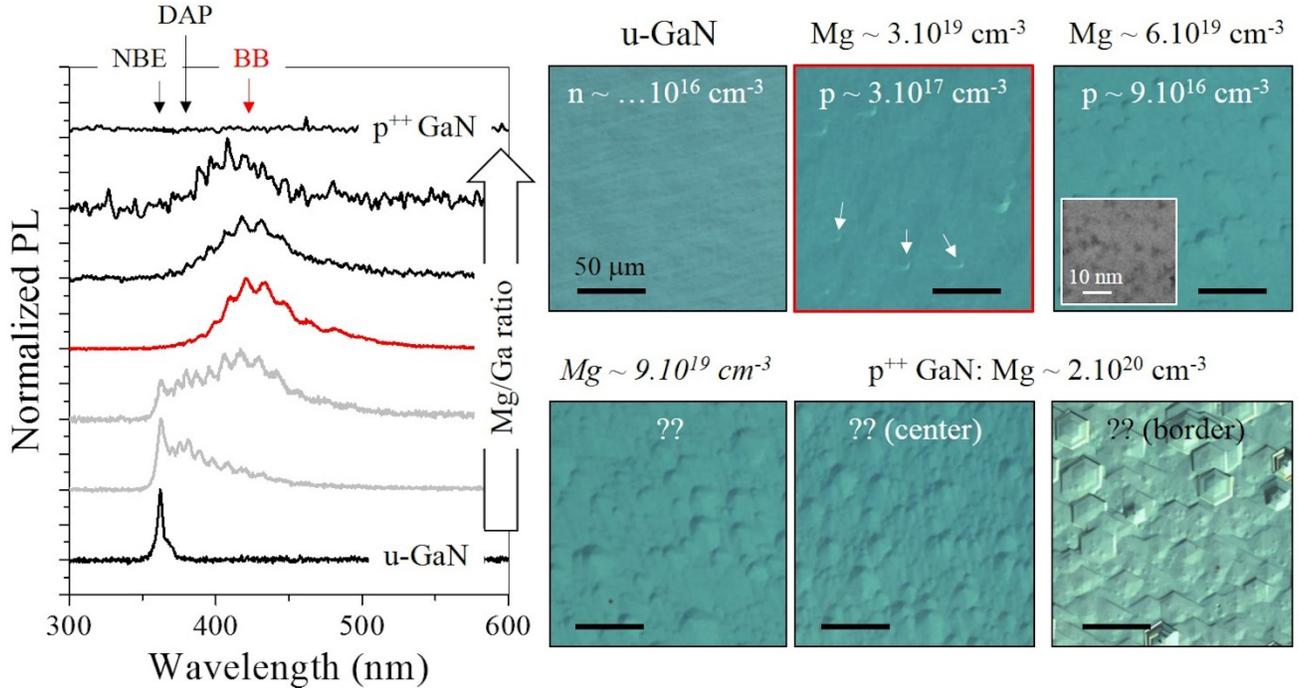

FIG. 1. Normalized PL emission spectra and optical microscope pictures of post-annealed GaN grown at 975°C at various Mg/Ga ratio. The inset shows the pyramidal IDs observed in cross-section by TEM.

Fig. 2 shows a similar study on 500 nm thick (crack-free) Si doped GaN layers grown at various Si/Ga ratio using TMSi. The electron concentration increases with the Si/Ga ratio both for growth temperature of 975 and 1100°C. At Si concentration exceeding $1.5\times10^{19}$ cm$^{-3}$, the surfaces show large scale-like islands of different heights separated by pitted boundaries. Further increase of the Si level leads to extremely rough surfaces characterized by larger pits and smaller grains, similar to previous reports [5,40]. Under typical MOVPE growth conditions and for a concentration higher than $10^{19}$ cm$^{-3}$, Si acts as an anti-surfactant. It was suspected to locally form Si-Ga-N substructures, which initiate a SiN$_x$ nano-masking that promotes 3D growth [41,42]. However, based on fine HR-TEM analysis, it is more likely a SiGaN$_3$ monolayer that inhibits the 2D growth [43]. Thus at the doping levels used for TJ application, i.e. above $10^{20}$ cm$^{-3}$, Si behaves as an anti-surfactant rather than a dopant, as it is incorporated inactively in SiGaN$_3$. We observed that the 2D-3D transition always happens around a doping level of $10^{19}$ cm$^{-3}$ whatever the growth temperature. Thus the 2D-3D transition is driven by the actual Si concentration on the surface and not by the surface kinetics of Ga or the precursor we used. The different Si/Ga ratio is due to the decomposition kinetics of the TMSi, whose decomposition temperature is estimated to be around 1000°C [44]. In other words, the supply of active Si, i.e. the effective Si/Ga ratio in the gas phase, depends with TMSi on the growth temperature. SiH$_4$, a more common precursor, does not present such disadvantage since it is fully decomposed around 700°C, and thus usually no temperature dependence of the roughening is reported.

Fig. 3 shows the SIMS profiles of two Si and Mg doped GaN layers grown successively at four different temperatures without varying the other growth parameters, i.e. keeping the NH$_3$, TMGa, TMSi and EtCp$_2$Mg flows constant. Unlike TMSi, the EtCp$_2$Mg decomposition and further Mg incorporation is stable over temperatures ranging from 975 to 1100°C. Mg piled up at the surfaces during the 5 min growth interruptions (GI) used to ramp up and stabilize the temperature between each step, which indicates we doped below the absolute solubility limit. Both for Si:GaN and Mg:GaN, the carbon level increases from $10^{16}$ to $10^{17}$ cm$^{-3}$ when the growth temperature decreases from 1100 to 975°C. This is commonly attributed to the lower decomposition rate of NH$_3$ and less active H readily available to react and etch away hydrocarbons species present on the surface and is also observed at similar level in undoped layers. Such levels do not significantly affect the GaN conductivity. Interestingly, the O signal is very different for both layers. While the background level remains below $10^{16}$ cm$^{-3}$ throughout the p-GaN growth, the O concentration increases of one order of magnitude when the n-GaN doping level reaches about $10^{19}$ cm$^{-3}$. Conversely, once the Si level decreases below $10^{19}$ cm$^{-3}$, the O concentration slowly returns to the background level. Since 3D growth starts above $10^{19}$ cm$^{-3}$ as mentioned previously, and 3D growth generates additional semi- and non-polar facets which are usually prone to higher O incorporation, the O signal correlates with the 2D-3D growth transition [45].

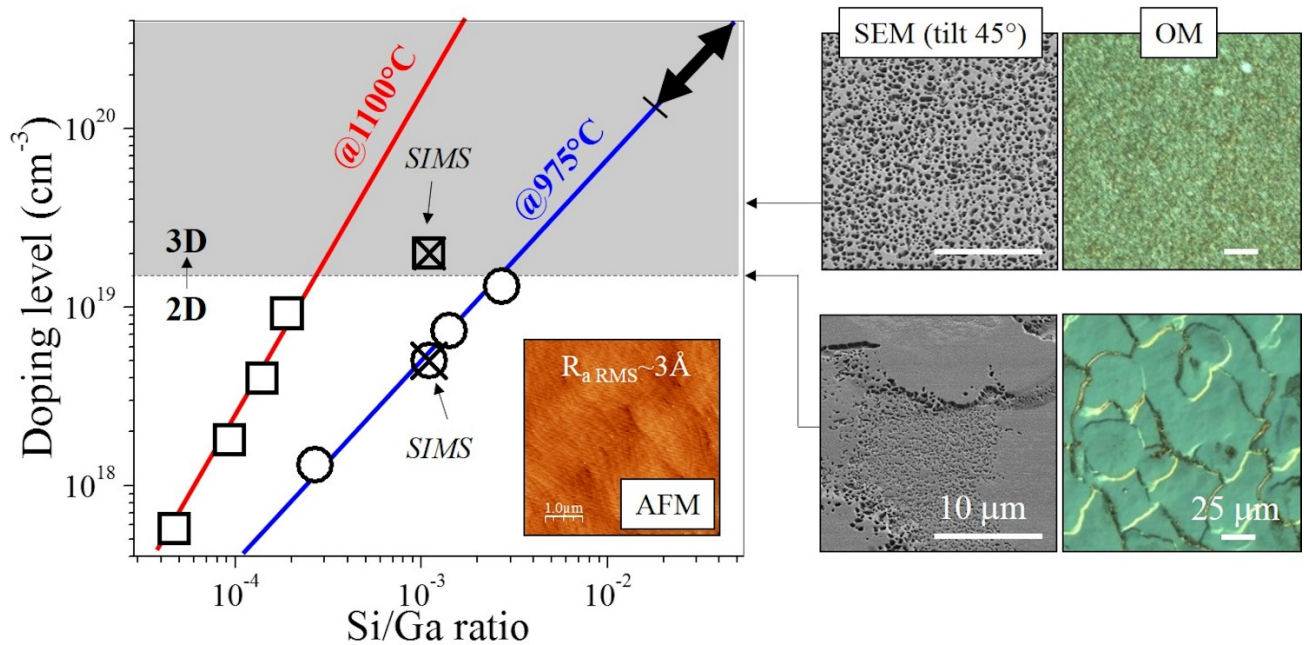

FIG. 2. Doping levels and surface morphology of Si:GaN layers grown at different Si/Ga ratio. A transition from 2D toward 3D growth occurs for Si concentration higher than $10^{19}$ cm$^{-3}$. The thick left-right arrow indicates the conditions typically required for the growth of TJ.

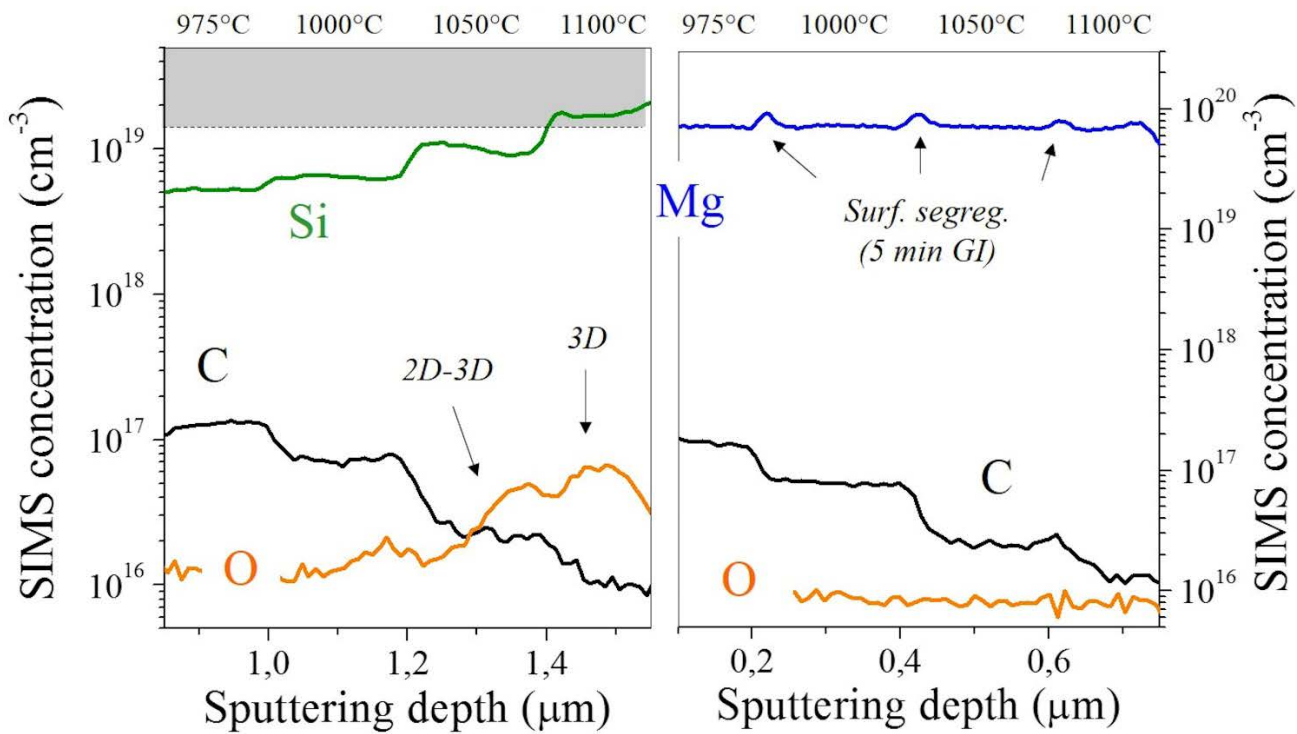

FIG. 3. SIMS analysis of n-GaN (left) and p-GaN (right) samples grown at different temperatures. The O signal increases of one order of magnitude when the Si level is higher than $10^{19}$/cm$^3$, i.e. when the growth mode changes from 2D to 3D. The Mg incorporation is constant over the full temperature range between 975 and 1100°C but Mg tends to accumulate on the surface during each growth interruption.

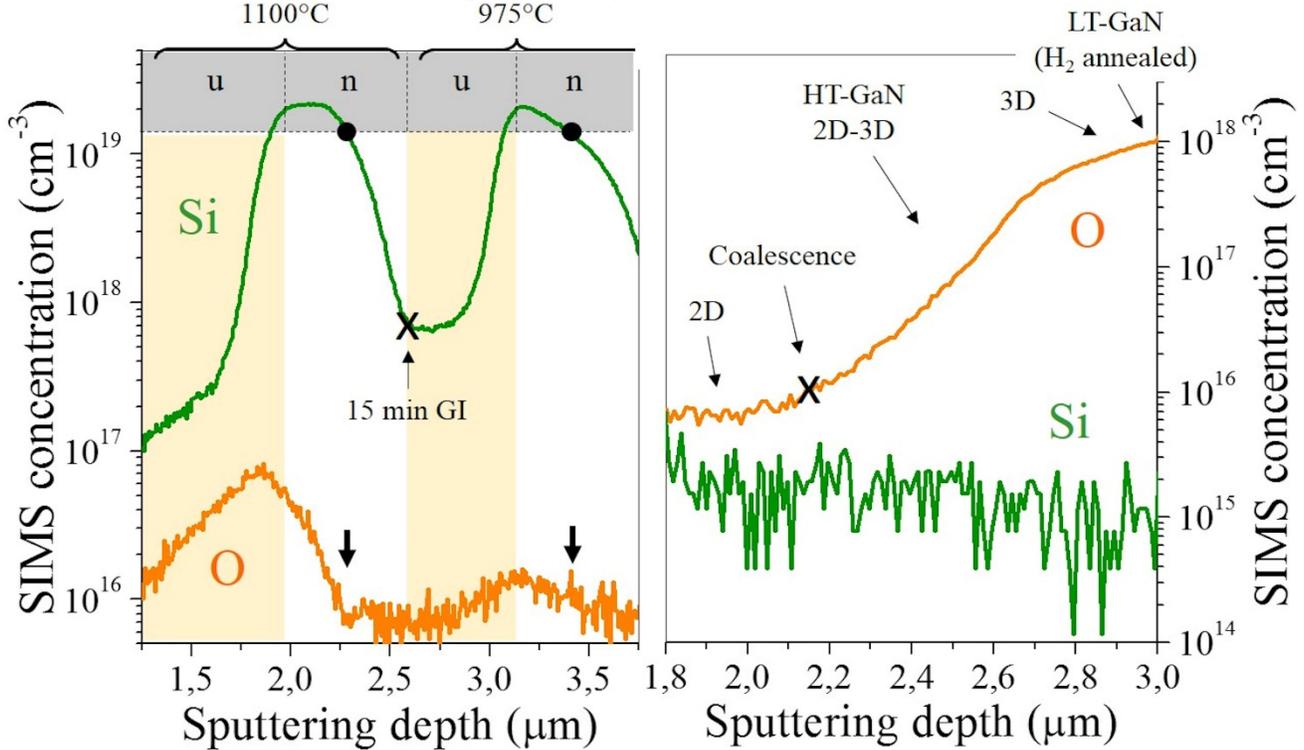

FIG. 4. SIMS analysis of n-GaN and u-GaN layers. An increase of the O level is observed each time the 3D growth mode is triggered, either by a Si concentration higher than $10^{19}$ cm$^{-3}$ (left) or a H$_2$ annealed LT buffer layer (right). Further GaN overgrowth of the rough surface with proper conditions leads to coalescence and the O level simultaneously decreases to below $10^{16}$ cm$^{-3}$.

Fig. 4 presents two additional evidences for correlation of the O level with 3D growth, triggered with and without SiN$_x$ nanomasking. For the first sample, 600 nm n-GaN is grown at 975°C with a linear increase of the Si concentration up to several $10^{19}$ cm$^{-3}$, followed by a 600 nm u-GaN to recover a smooth surface. After a 15 min growth interruption, the same sequence is repeated at 1100°C. It is worth noting that the O concentration increases as soon as the 3D growth is triggered by the SiN$_x$ nanomasking (black dots and arrows on Fig. 4 (left)). Conversely, the O level slowly decreases upon u-GaN overgrowth which again smooths the surface. The second sample shows a similar behavior (Fig. 4 (right)) during a u-GaN template growth on sapphire. Here, intentionally, 3D islands are obtained by annealing a nucleation layer deposited at low temperature (LT). At this point, the O level is around $10^{18}$ cm$^{-3}$. Further regrowth of the 3D nuclei at high temperature (HT) smooths the surface. Since this means that the semipolar facettes are vanishing, the O concentration also decreases until it reaches its background level of about $10^{16}$ cm$^{-3}$ for O incorporation on a flat (0001) surface.

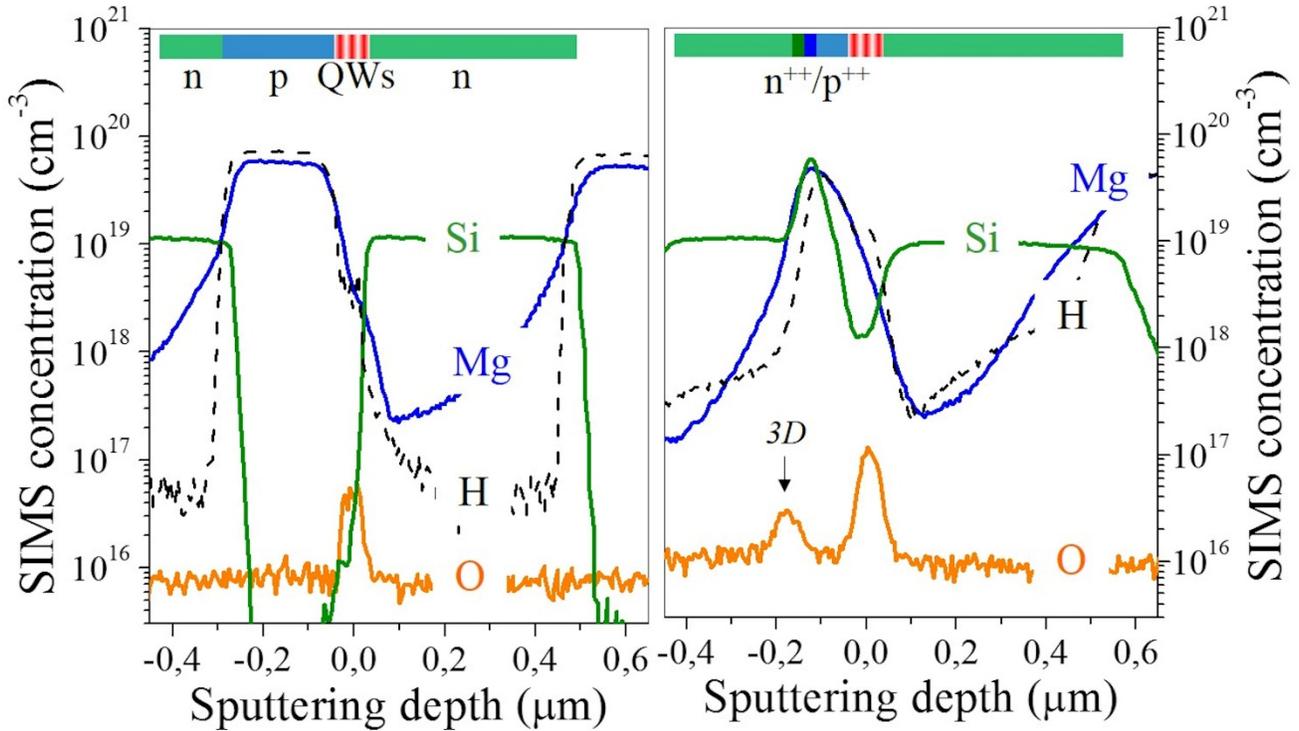

FIG. 5. SIMS analysis of the np-n (left) and TJ (right) based LEDs. Note the O peak and the Mg tail buried in the n$^{++}$ side of the TJ. The x-axis origin has been set at the QWs position for better comparison.

Fig. 1, 2, 3 and 4 indicate a rather low doping limit on both Si:GaN and Mg:GaN, which is theoretically unsuitable for a TJ. However, driven by many successful reports found in the literature, we compared the SIMS profile of a np-n LED (no n$^{++}$/p$^{++}$ layers at the p-n junction) and a TJ based LED. Fig. 5 (left) shows the results of the np-n LED structure featuring the highest doping level as determined from Fig. 1 and 2 (~$10^{19}$ and few $10^{19}$ cm$^{-3}$ for Si and Mg respectively). The O level exhibits a peak around the QWs position, most likely due to the residual O found in the N$_2$ carrier gas required for the InGaN growth (n- and p-GaN are grown under pure H$_2$ carrier gas, which is easier to purify and present lower contamination levels). Since Mg has a strong carry-over into subsequent layers, it is also incorporated in the Si:GaN top layer [46-48]. Even though Mg decreases exponentially, the first ten nanometers of Si:GaN have still a higher Mg than Si concentration and hence are rather insulating. Important to notice is that the Mg is potentially active in this region since the hydrogen level correlates with Mg, and hydrogen does not incorporate well in Si:GaN [49]. In conclusion, the electrical interface of the maximum doped np top junction is not as well defined as intended. For comparison, we realized the same structure with a 15 nm/15 nm n$^{++}$/p$^{++}$ TJ

inserted between the n- and p-GaN top layers. The Si and Mg levels were aimed to be around $2\times10^{20}$ and $1\times10^{20}$ cm$^{-3}$ respectively. The SIMS analysis in Fig. 5 (right) shows that an additional O peak appears in the n$^{++}$ region, indicating a rough interface due to 3D growth. Unfortunately, the SIMS profiles lack depth resolution, i.e. the concentration of the different elements is averaged over a larger volume, depending on the shape of the sputter crater at this position. This underestimates the concentration in these layers. Still, when reaching appropriate Si:GaN conditions, the rough surface can be smoothened by a proper overgrowth. Thus, as expected from Fig. 2, the 15 nm thin n$^{++}$ layer has a lot of nanomasking, is very rough, and likely discontinuous, i.e. it may be even insulating in some places, in short not appropriate for an abrupt TJ.

Fig. 6 (left) compares the IV characteristics of three TJ based LEDs with different p$^{++}$ and n$^{++}$ doping levels. The reference is a standard cyan LED emitting at a wavelength of 490 nm (red curve). In good agreement with the results found in the literature, the TJ based LEDs show typical LED behavior and emit light once both Mg and Si expected levels reach about $10^{20}$ cm$^{-3}$ [31]. A further increase of either dopant leads to a significantly decrease of the penalty voltage. Si seems more efficient than Mg at lowering the junction barrier, as previously observed by Kaga *et al.* [32]. Eventually, for Mg and Si levels of $10^{20}$ and $2\times10^{20}$ cm$^{-3}$ respectively, the TJ based LED penalty voltage is reduced to as low as 1.1 V with a specific differential resistance of about $10^{-2}$ $\Omega.\text{cm}^2$ at 20 mA. These values are surprisingly low considering the present structure does not involve any (p-)InGaN interlayer that is often introduced between the p$^{++}$ and n$^{++}$ layers to reduce the nominal tunnel barrier height and facilitate the tunneling process [29,50].

Based on the rather poor morphological, crystalline, and electric properties of the overdoped doped n$^{++}$- and p$^{++}$-GaN layers presented in the previous sections, the results of Fig. 6 (left) seriously challenge a pure tunneling mechanism for the following reasons:

i) The maximum free carrier density achievable in the p$^{++}$ and n$^{++}$ side of the junction is limited around $3\times10^{19}$ and $2\times10^{19}$ cm$^{-3}$ respectively (assuming a complete ionization of both dopants). Therefore, the space charge region (SCR) spreads over a distance of 20 nm which is not compatible with a Fowler-Nordheim emission, given the large bandgap of GaN.

ii) The Mg carry-over causes fairly high amount of active, i.e. not passivated, Mg ($>10^{19}$ cm$^{-3}$) to be incorporated in the n$^{++}$ side. This leads to a significant compensation of the free electron density, especially at the vicinity of the interface, which widens the SCR and so further reduces the electric field across the junction.

iii) Due to the effective doping asymmetry p$^{++}$>n$^{++}$ deduced from i) and ii), the SCR is shifted inside the n$^{++}$ layer. This part of the junction is characterized by a low structural quality and a high O contamination at grain boundaries inherited from a SiN$_x$ induced 3D growth. Worse, the presence of Mg likely induces additional defects such as IDs, or Mg-O and Mg-Si complexes. Thus, the field of the SCR is strongly affected by (various) deep centers that prevent the free carriers to directly tunnel from one side to the other one, and could even pin the Fermi level at their midgap states.

iv) Any attempt to increase the Mg and Si concentrations up to $10^{20}$ cm$^{-3}$ exacerbates the problems discussed in points ii) and iii) even though the penalty voltage decreases.

These considerations imply that a trap-assisted tunneling, i.e. a Poole-Frenkel emission, is the main mechanism responsible for the TJs in GaN [51]. The band diagram of such TJ and the proposed tunneling mechanisms are sketched in Fig. 6 (right). Instead of a direct band to band tunneling, the carriers tunnel through defect states. Thus a less abrupt band transition can still allow relatively efficient tunneling. The tunnel current is then rather limited by the distance of the traps, and a more trap rich material could sustain higher tunnel current. It is worth pointing out a similar mechanism should occur with AlGaN. Here the problems are even worse since the larger gap would require even higher dopant concentrations and associated issues. In addition, at higher levels, the Si forms DX centers in AlGaN. Hence, overdoping would lead to more trap states and explain the recently reported relatively high conductivity of AlGaN TJ in deep UV LEDs [21].

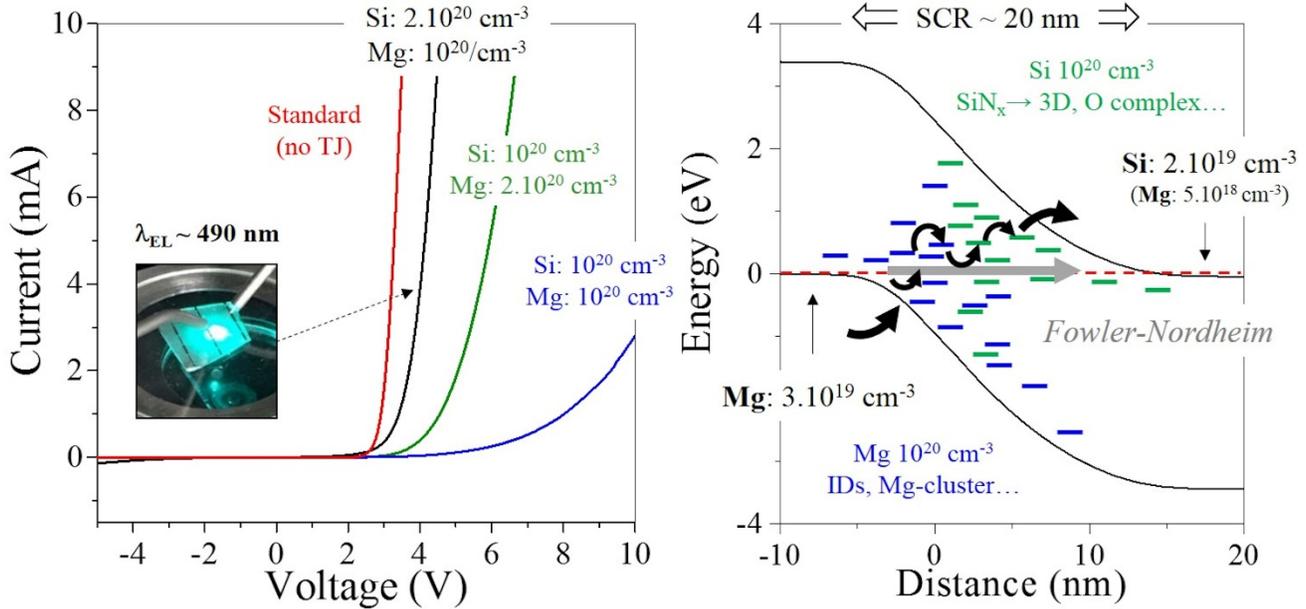

FIG. 6. I-V characteristics of standard and TJ based LEDs with different $p^{++}$ and $n^{++}$ doping levels (left). Band diagram of the unbiased TJ assuming that the p and n doping levels saturate at $3\times10^{19}$ and $2\times10^{19}$ cm$^{-3}$ respectively, and taking into account the carry-over of Mg (right). The curved black arrows sketches the trap-assisted tunneling (Poole-Frenkel emission). The straight grey arrow represents a pure tunneling (Fowler-Nordheim emission). Note the position of the defect levels is arbitrary.

We must stress that our results are obtained by MOVPE growth. A TJ grown by molecular beam epitaxy (MBE) does not significantly suffer from Mg carry-over and Mg-H passivation, even when using NH$_3$ as nitrogen precursor [52]. In addition, higher doping levels for smooth layers have been achieved by MBE up to a few $10^{20}$ cm$^{-3}$ [53,54].

Therefore, it seems interesting to discuss TJs found in the literature in the light of our observations and results. For instance, Young *et al.* observed the TJ formed by MBE overgrowth of a MOVPE grown p-n LED by atomic probe tomography (APT). They observed a high O concentration at the TJ interface, most likely originating from the residual Ga$_x$O$_y$ formed during the sample transfer from the MOVPE reactor to the NH$_3$-MBE chamber [55]. In contrast to our conclusion that O introduces midgap states in the SCR, the authors concluded the O rather establish a delta doping that strongly reduces the SCR. However, regrowth on a partly Ga$_x$O$_y$ covered template can easily lead to 3D growth due to different attachment rates on GaN and Ga$_x$O$_y$. More recently, Akatsuka *et al.* reported that an overlap of Mg and Si doping at the TJ interface enhances the tunneling current in MOVPE grown

TJs [56]. This is counterintuitive to a sharp TJ, so the authors speculated that either GaN bandgap narrowing due to heavy doping or an advanced "2Mg+1Si" p-GaN co-doping (proposed by Katayama-Yoshida *et al.* [57]) would explain the lower resistivity of their device. However, their results are very reasonable when assuming that traps are needed to enhance tunneling. Indeed, the strong crystal disorder introduced at the TJ interface most likely favors the creation of traps that can assist the tunneling of the carriers. A similar argument applies to the so-called polarization enhancement TJ structures. Theoretically, the insertion of a thin InGaN interlayer between the $p^{++}$ and $n^{++}$ GaN allows a narrower SCR (due to the local bandgap reduction) and a higher electric field across the junction (due to built-in spontaneous and piezoelectric polarization), which is beneficial for carrier tunneling [58]. Practically, one should also consider that InGaN is grown at low temperature, and thus has a much higher defect density than GaN. In the case of MOVPE, InGaN must be grown under $N_2$ ambient and features a much higher O background (as seen in Fig. 5) while high quality n- and p-GaN require $H_2$ and higher temperature. Thus, one has to choose between growing the whole TJ at same temperature under $N_2$ (resulting in poorer crystal quality and higher background O levels) or interrupting the growth to properly switch the carrier gas and increase the temperature (leading to Mg and eventual contaminations piling up at each interface). Furthermore, In-rich InGaN easily introduces 3D growth after a few nanometers [59].

Whatever approach is chosen, either overdoping, MBE regrowth or InGaN interlayers, all of these introduce a fairly large amount of disorder inside the TJ, leading to a tunneling current mainly supported by the defects. Following this assumption, one may enhance the tunneling current by deliberately introducing deep levels (e.g. doping with carbon, europium, or iron…).

## IV. CONCLUSION

We systematically studied the electrical properties, the surface morphology and the Mg, Si, H, C and O concentrations of p-GaN and n-GaN layers grown by MOVPE at different temperatures for dopant levels ranging from $10^{18}$ to $10^{20}$ cm$^{-3}$. SIMS results show the C concentration drastically increases from $10^{16}$ cm$^{-3}$ at 1100°C to about $10^{17}$ cm$^{-3}$ at 975°C without significantly affecting the electrical properties of both p- and n-type layers. However, O is drastically increased once the Si concentration exceeds $10^{19}$ cm$^{-3}$ and the growth mode simultaneously changes from 2D to 3D due to $SiN_x$ nanomasking. Overdoping with Mg to $10^{20}$ cm$^{-3}$ causes inversion domains and defect rich layers which show no longer PL. Thus, at the high Si/Ga and Mg/Ga ratios typically used for $p^+$-$n^+$ tunnel junctions, both type of layers are resistive and present rough surfaces or inversion domains. SIMS analysis of a np-n LED indicates a strong Mg carry-over. The pinning of the Fermi level, close to the conduction band in the n-GaN, destabilizes the Mg-H complex (Mg/H<<1) and allows the unintentionally incorporated Mg to efficiently compensate the doping level of the first nanometers of the n-GaN top layer. However, introducing more defective $n^{++}$ and $p^{++}$ layers lead to the desired tunneling. Increasing dopant concentration (and thus defects) reduces the penalty voltage. Based on these results, we conclude that tunnel junctions grown by MOVPE present a poor structural quality and mostly work in defects assisted tunneling.

## ACKNOWLEDGMENTS

Y.R. would like to thank T. Takeuchi from Meijo University and Z. Sitar from North Carolina State University for fruitful discussions on p-GaN, n-GaN and TJ grown by MOVPE. Interesting discussions with M. Leroux, B. Damilano and F. Semond from CRHEA-CNRS about growth and properties of (doped) nitride layers made by MBE were particularly appreciated. This work was supported by Aichi Science and Technology Foundation Knowledge Hub Aichi Priority Research Project E and by JST (Japan Science and Technology Agency), Strategic International Collaborative Research Program, SICORP.

------------------------